\documentclass[a4paper,11pt]{article}
\pdfoutput=1 

\usepackage{jcappub} 
\usepackage{caption} 
\title{\boldmath Dark matter subhalos and unidentified sources in the Fermi 3FGL source catalog}

\date{\today}

\author[a,b]{Djoeke Schoonenberg,}
\author[b]{Jennifer Gaskins,}
\author[b]{Gianfranco Bertone,}
\author[c]{and J{\"u}rg Diemand}
\affiliation[a]{Anton Pannekoek Institute, University of Amsterdam, The Netherlands}
\affiliation[b]{GRAPPA, University of Amsterdam, The Netherlands}
\affiliation[c]{Institute for Computational Sciences, University of Z{\"u}rich, Switzerland}
\emailAdd{d.schoonenberg@uva.nl}
\emailAdd{jgaskins@uva.nl}
\abstract{
If dark matter consists of weakly interacting massive particles (WIMPs), dark matter subhalos in the Milky Way could be detectable as gamma-ray point sources due to WIMP annihilation. In this work, we perform an updated study of the detectability of dark matter subhalos as gamma-ray sources with the Fermi Large Area Telescope (Fermi LAT).
We use the results of the Via Lactea II simulation, scaled to the Planck 2015 cosmological parameters, to predict the local dark matter subhalo distribution.  Under optimistic assumptions for the WIMP parameters --- a 40 GeV particle annihilating to $b\bar{b}$ with a thermal cross-section, as required to explain the Galactic center GeV excess --- we predict that at most $\sim 10$ subhalos might be present in the third Fermi LAT source catalog (3FGL).  This is a smaller number than has been predicted by prior studies, and we discuss the origin of this difference. 
We also compare our predictions for the detectability of subhalos with the number of subhalo candidate sources in 3FGL, and derive upper limits on the WIMP annihilation cross-section as a function of the particle mass. If a dark matter interpretation could be excluded for all 3FGL sources, our constraints would be competitive with those found by indirect searches using other targets, such as known Milky Way satellite galaxies.
}
\begin{document}

\maketitle

\section{Introduction}
About 85\% of all matter in the Universe is dark matter \cite{Bertone:2010zza,Jungman:1995df,Bergstrom00,Bertone05}.  
The most popular class of dark matter particle candidates is that of \emph{weakly interacting massive particles} (WIMPs). These particles achieve the appropriate relic density through self-annihilation in the early-universe, and can be searched for with astrophysical data through the measurement of secondary particles produced in their annihilation in regions of high dark matter density (see e.g. Refs. \cite{Bringmann:2012ez,Klasen:2015uma} for recent reviews), including the Galactic center (see e.g. Refs. \cite{Berezinsky:1994wva,Gondolo:1999ef,Bertone:2002je,Merritt:2002vj,Cesarini:2003nr,Hisano:2004ds,Horns:2004bk,Aloisio:2004hy,Bertone:2005hw,Merritt:2006mt,Dodelson:2007gd,Regis:2008ij,Serpico:2008ga}), subhalos in the Milky Way  \cite{Silk:1992bh,Berezinsky:1996eg,Stoehr:2003hf,Diemand:2006ik,bergstrom,pieri,Pieri:2007ir, SiegalGaskins:2007dx, kuhlen, Anderson:2010df, Buckley:2010vg, zechlin, Zechlin:2011kk, Belikov:2011pu, Mirabal:2012em, berlinhooper, bertoni} and dark matter mini-spikes around intermediate-mass black holes \cite{Zhao:2005zr,Bertone:2005xz,Horiuchi:2006de,Fornasa:2007ap,Brun:2007tn,Taoso:2008qz,Bringmann:2009ip,Moulin:2009zza,Bertone:2009kj,Sandick:2010yd,Eda:2013gg,Eda:2014kra,Wanders:2014xia}. 

In this study we perform an updated analysis of the detectability of point-like dark matter subhalos with the Fermi Large Area Telescope (Fermi LAT).  Taking the host halo of the Via Lactea II (VL-II) simulation as a proxy for the dark matter halo hosting the Milky Way, we calculate the number of dark matter subhalos one would expect to appear as unidentified sources in the recently released Fermi LAT third source catalog (3FGL) \cite{fermi3fgl} for different benchmark dark matter particle models. 
We consider the actual subhalos present in the VL-II simulation, with halo concentrations scaled to the latest cosmological parameters as measured by Planck \cite{planck2015}, to determine the local subhalo distribution. In this article by subhalos we mean all halos within 400 kpc from the Galactic center (roughly the virial radius). We include all subhalos in VL-II that are well resolved in the VL-II simulation, i.e. those more massive than $\sim 10^5 M_{\odot}$.
We compare our predicted number of detectable subhalos to the number of dark matter subhalo candidate sources in 3FGL  identified in Ref. \cite{bertoni} to place upper limits on the dark matter annihilation cross-section. Our work updates the predictions of Ref. \cite{kuhlen}, who adopted a similar approach using VL-II, and compares those predictions to the most current source catalog available from the Fermi LAT\@.

Our analysis also provides a test of the dark matter interpretation of the gamma-ray excess in data from the Fermi Large Area telescope (LAT) that has been recently discovered from the center of our Galaxy ~\cite{goodenough, Vitale:2009hr, Hooper:2011ti, Gordon:2013vta, Abazajian:2014fta, Daylan:2014rsa, Calore:2014xka, fermigc}. Although astrophysical sources might explain the excess \cite{Abazajian:2010zy,Calore:2014oga,Yuan:2014rca, Petrovic:2014xra,Carlson:2014cwa,Petrovic:2014uda,Cholis:2015dea,Gaggero:2015nsa,richard,Lee:2015fea,Calore:2015bsx}, and dwarf galaxies partially constrain the dark matter interpretation, we provide here complementary constraints by estimating the number of subhalos that should have been detected by Fermi LAT assuming values of the WIMP mass and annihilation cross-section compatible with the Galactic Center excess.

The paper is organised as follows: in section \ref{subhalos}, we discuss the subhalos in VL-II that we use as a proxy for the subhalos in the Milky Way and their WIMP annihilation flux. In section \ref{3fglsection}, we discuss the 3FGL source catalog of Fermi LAT and the energy flux thresholds we use to define subhalo detectability. Our results are presented in section \ref{results-section}. Finally, a discussion of the results and our conclusions are given in section \ref{discussion-section}.

\section{Gamma rays from subhalos}
\label{subhalos}
\subsection{Subhalos in Via Lactea II}
In the past decade, cosmological numerical simulations run on powerful supercomputers led to a substantial improvement in our capability to estimate the dark matter distribution in galaxies such as the Milky Way. We make use in this paper of the Via Lactea II (VL-II) simulation \cite{vl2}, which includes only dark matter particles and does not take into account baryonic effects.
The simulated MW halo contains numerous smaller substructures, as predicted by the $\Lambda$CDM theory of hierarchical structure formation. The smallest subhalos resolved in VL-II are $\sim 10^{5} M_{\odot}$.
We discard all subhalos at a distance greater than 400 kpc (approximately the virial radius of the Milky Way) from the Galactic center, since small halos (compared to the host mass of $\sim 10^{12} M_{\odot}$) at larger radii are unlikely to produce interesting results.

In this work, we model the density profiles of the VL-II subhalos using a Navarro-Frenk-White (NFW)  \cite{NFW} profile
\begin{equation}\label{nfw1}
\rho(r) = \frac{\rho_{s}}{\left( \frac{r}{r_{s}} \right) \left[ 1 + \left( \frac{r}{r_{s}} \right) \right]^{2}}
\end{equation}
where $\rho_{s}$ and $r_{s}$, the scale density and scale radius, respectively, are characteristics of individual halos. Assuming an Einasto profile instead of a NFW profile would result in only marginally higher luminosities \cite{pieri}. 

When a subhalo falls into the host halo, tidal stripping removes mass from the outer parts of subhalos \cite{diemand}. This effect can be approximated by removing the mass beyond a `tidal' radius, defined as the radius at which the subhalo density is equal to the host density. Since the tidal radius is often larger than the NFW {\it scale radius} of an infalling halo, the matter within the NFW scale radius of a halo is mostly preserved \cite{tidalstripping}. Indeed, in VL-II, the tidal radii of halos (defined as the radii at which the subhalo densities equal the host halo density) are always larger than their scale radii.

In order to calculate the NFW parameters, we extract from the numerical simulation $V_{max}$ and $r_{Vmax}$, defined as the maximum circular velocity of the subhalo, and the radius at which this velocity is reached.  We scale $r_{\text{Vmax}}$ to the Planck 2015 cosmological parameters (see Appendix). To be conservative, we do not include the possible annihilation boost due to sub-substructures, i.e. substructures within subhalos (e.g., \cite{Anderson:2010df,sanchezconde}).

\subsection{Gamma rays from WIMP annihilations}\label{g-rays}

In this work, we consider the continuous gamma-ray spectrum produced as a result of annihilations of (Majorana) WIMPs in dark matter subhalos.
The differential annihilation flux in photons per unit time, area, and energy from a dark matter halo is given by
\begin{equation}\label{fluxje}
\frac{d \Psi}{d E_{\gamma}} = \frac{d N}{d E_{\gamma}} \frac{\langle \sigma v\rangle}{8 \pi m_{\chi}^{2}} \cdot J,
\end{equation}
where $ \frac{d N}{d E_{\gamma}} $ is the number of photons per photon energy produced per annihilation, $\langle \sigma v \rangle$ is the relative velocity of the particles times the annihilation cross-section, averaged over the velocity distribution, $m_{\chi} $ is the dark matter particle mass, and $J$ (the ``astrophysical'' term) is given by
\begin{equation}\label{2}
J = \int_{\Omega} d \Omega \int_{l.o.s.} \rho_{\chi}^{2} (l, \Omega) dl
\end{equation}
where $\rho_{\chi}$ is the dark matter mass density, and the integration is along the line-of-sight $l$ and over the solid angle $\Omega$. 
We obtain the gamma-ray spectrum $dN/dE_{\gamma}$ in Eq.~\ref{fluxje} from \cite{PPPC}.

If one considers the object of interest (e.g., a dark matter subhalo) to be at a distance great enough for it to be treated as a point source, one can write
\begin{equation}\label{jpointlike}
J_{\text{pointlike}} = \frac{1}{d^{2}} \int_{0}^{r_{s}} 4 \pi r^{2} \rho(r)^{2} dr
\end{equation}
where we choose to integrate out to the scale radius instead of the virial radius, since \hbox{$\sim 90$ \%}\break of the luminosity is generated within the scale radius. 

For simplicity, we are interested in subhalos that would appear point-like to Fermi LAT, since determining the detectability of an extended source with the Fermi LAT depends strongly on the location in the sky and the properties of nearby sources and diffuse emission.
For this reason, we declare a subhalo to be spatially extended if its angular extent on the sky is larger than the 68 \% containment angle of Fermi LAT at 1 GeV, since the gamma-ray spectrum resulting from WIMP annihilations peaks around this energy (see section \ref{g-rays}).
The 68 \% containment angle at 1 GeV of the Fermi LAT is $\sim 0.8$ degrees for Pass 7 source events (P7REP\_SOURCE\_V15), used for the production of 3FGL \cite{fermi3fgl}.
Thus, our spatial extension threshold for defining a source as point-like is
\begin{equation}\label{spatialextension}
\text{ext} = \frac{180}{\pi} \arctan(r_{s} / d) < 0.8 \: \text{degrees}
\end{equation}
where $r_{s}$ is the scale radius and $d$ is the distance between the observer and the subhalo. The $J$-factors of point-like subhalos were calculated using Eq. \ref{jpointlike}. For spatially extended subhalos, however, we will correct the $J$-factors for spatial extension by integrating the luminosity out to the containment angle rather than the scale radius. The results for extended sources will be presented separately.

\begin{figure}[t]
	\centering
		\includegraphics[width=0.45\textwidth]{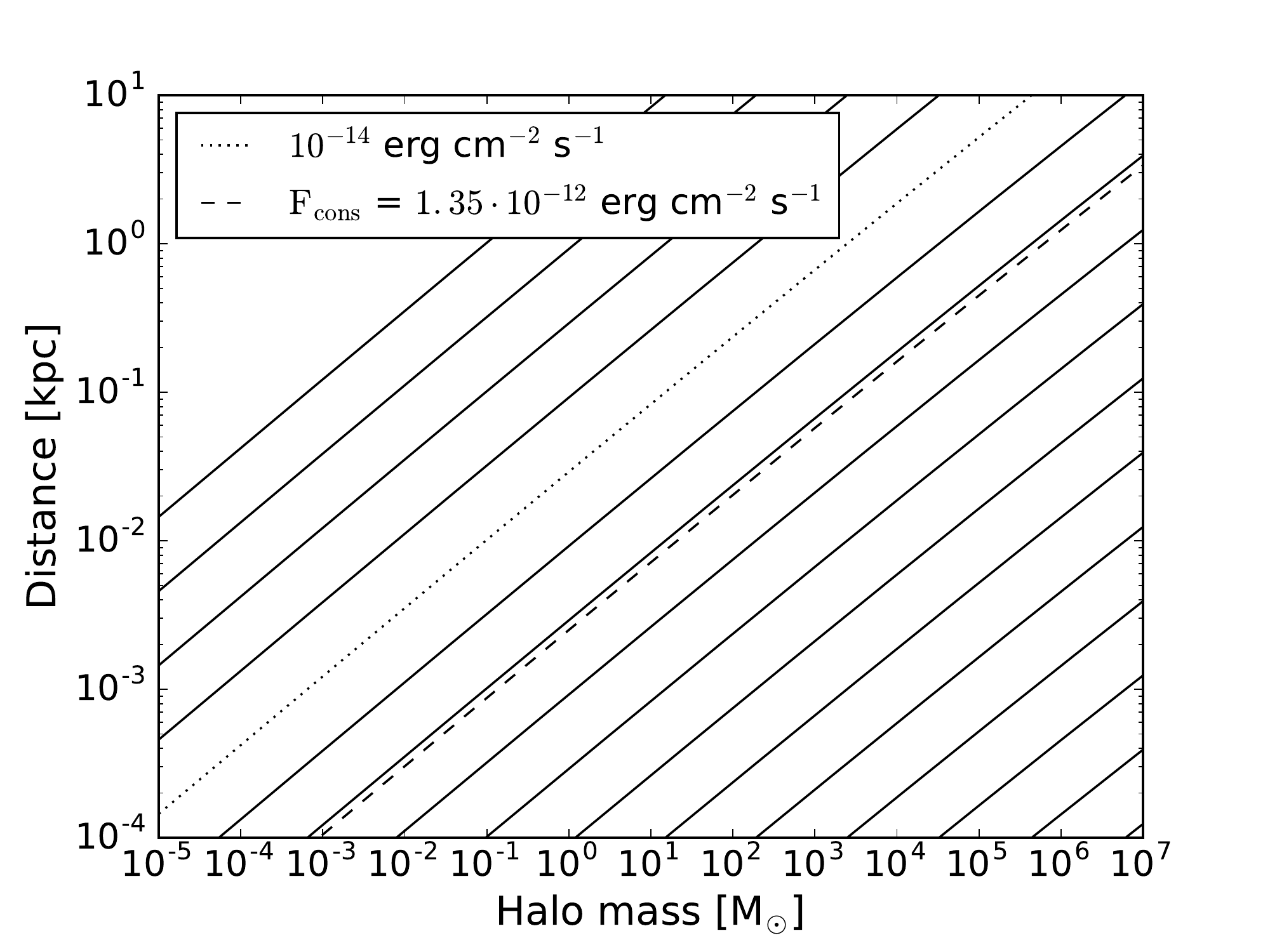}
		\includegraphics[width=0.45\textwidth]{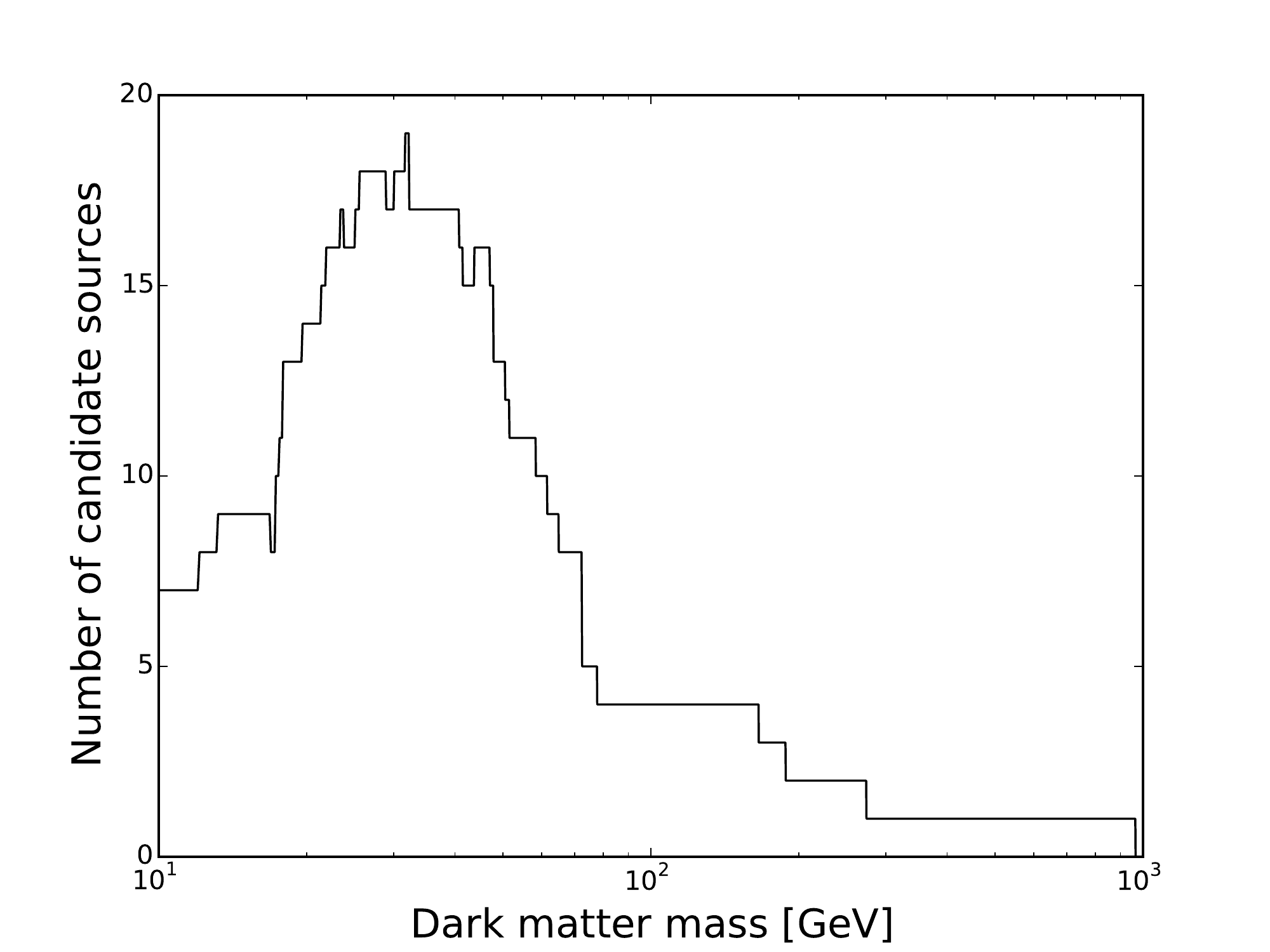}
\caption{\emph{Left:} Contours of constant gamma-ray flux in the energy range 1 $-$ 100 GeV from dark matter subhalos of given mass as a function of their distance from the observer, assuming a 100 GeV dark matter particle annihilating to $b\bar{b}$ at a thermal cross-section. Moving one contour to the right corresponds to an increase in flux of one order of magnitude, with the dotted line indicating a flux of \hbox{$10^{-14}$ erg $\text{cm}^{-2} \: \text{s}^{-1}$} for reference. The dashed line corresponds to the conservative flux threshold for inclusion in 3FGL used in the rest of this paper (see section \ref{3fglsection}). \emph{Right:} The number of sources in 3FGL identified as possible dark matter subhalos in Ref. \cite{bertoni} as a function of dark matter particle mass, in the case of annihilations to $b\bar{b}$.\label{mdisflux_NFW}}
\end{figure}

In the left panel of Fig.~\ref{mdisflux_NFW} we show contours of constant flux in the distance vs.~dark matter halo mass plane. The $J$-factors of subhalos were calculated using Eq. \ref{jpointlike} and assuming NFW profiles with concentrations given by Eq.~11 in \cite{pieri} at $R_{GC} = 8$ kpc. The corresponding fluxes were calculated for the case of a 100 GeV dark matter particle annihilating to $b\bar{b}$ at a thermal cross-section $\langle\sigma v\rangle \: = 3 \cdot 10^{-26} \: \text{cm}^{3} \: \text{s}^{-1}$. Note that for a photon flux threshold of $1.35 \cdot 10^{-12}$ erg $\text{cm}^{-2} \: \text{s}^{-1}$ (dashed line in the figure), which corresponds to the peak of the distribution of energy fluxes above 1 GeV of sources in 3FGL (see left panel of Fig.~\ref{efluxhistscombined} and section \ref{3fglsection}), subhalos with masses less than 10$^7$ M$_\odot$ are not detectable at distances greater than $\sim 3$~kpc. 

\section{The 3FGL source catalog}\label{3fglsection}
Recently, the Fermi LAT collaboration has released the 3FGL catalog \cite{fermi3fgl}, which includes sources that were detected at 5$\sigma$ with 6 years of data.  In total, this catalog contains 3034 sources, many of which are associated with Active Galactic Nuclei or pulsars, thanks to astronomical studies at other wavelengths. There are 992 sources, however, that have not been associated with emission at other wavelengths. A subset of these are expected to be pulsars and AGN that have {\it not yet} been associated.  Sources that were not associated in earlier Fermi LAT catalogs have subsequently been associated with pulsars or AGN in 3FGL, and the same is expected to happen with more multi-wavelength data for many of the unassociated sources in 3FGL (see, e.g., the analysis of unassociated 2FGL sources in \cite{Mirabal:2012em}). However, there remains the possibility that a subset of the unassociated sources might be dark matter subhalos.

We will compare our predicted number of detectable subhalos to the number of {\it candidate} dark matter subhalos in 3FGL as found by \citet{bertoni}. Discarding time-variable unidentified sources, they performed spectral tests to determine the compatibility of the spectra of the unidentified sources with WIMP annihilation spectra. For unidentified sources which they identified as spectrally compatible with dark matter subhalos, they provided dark matter mass ranges for which the annihilation spectra provide good fits to the data in the case of annihilations to $b\bar{b}$, based on a chi-squared analysis.
The number of compatible unidentified sources as a function of dark matter mass, as found by \cite{bertoni}, is shown in the right panel of Fig. \ref{mdisflux_NFW}.
The relatively large number of candidate sources corresponding to dark matter masses of 20 $-$ 40~GeV are likely pulsars, as typical gamma-ray pulsar spectra are similar to those for a dark matter particle in that mass range annihilating to $b\bar{b}$~\cite{Baltz:2006sv}. 

\begin{figure*}[t]
	\centering
		\includegraphics[width=0.45\textwidth]{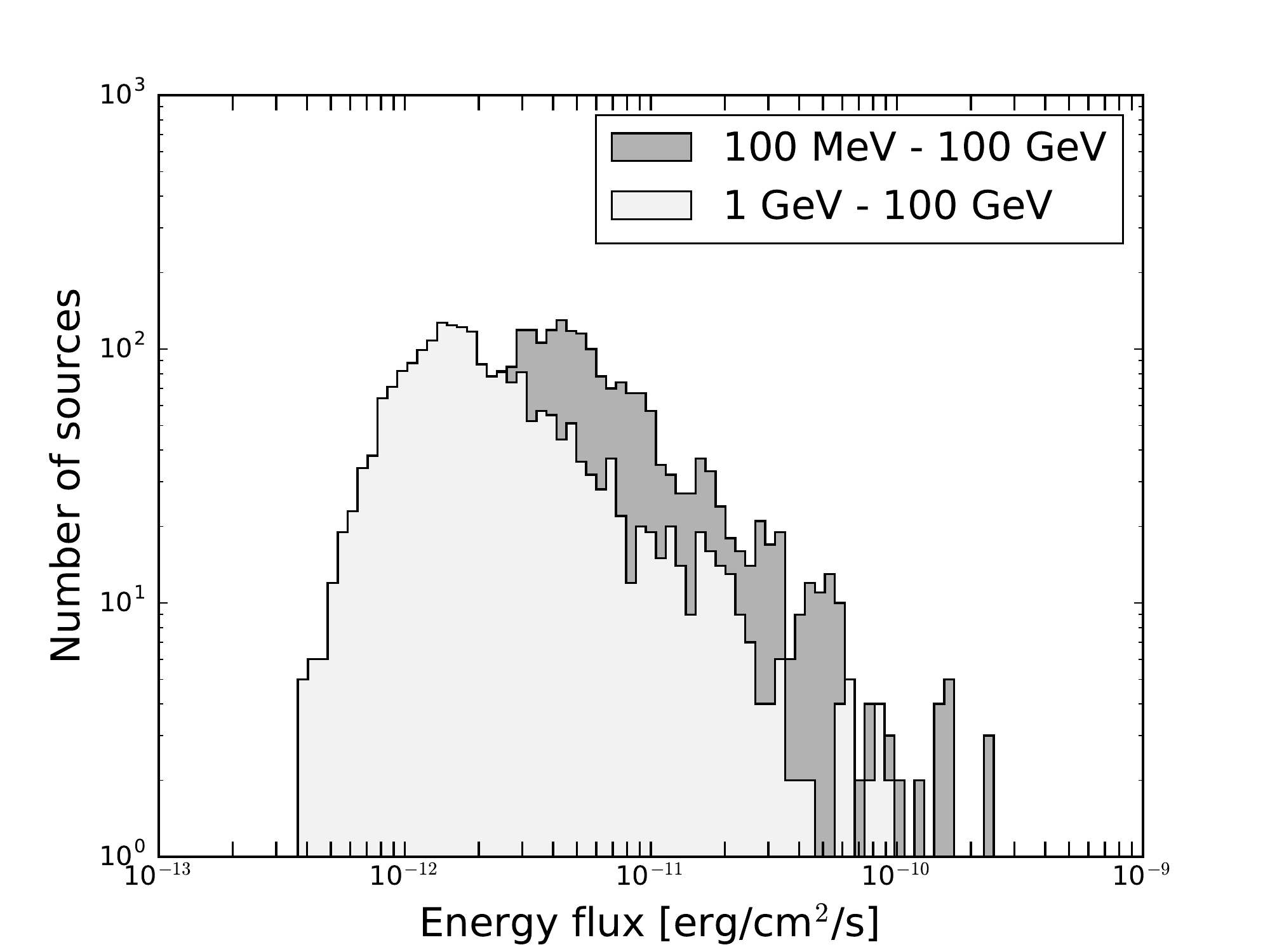}
		\includegraphics[width=0.45\textwidth]{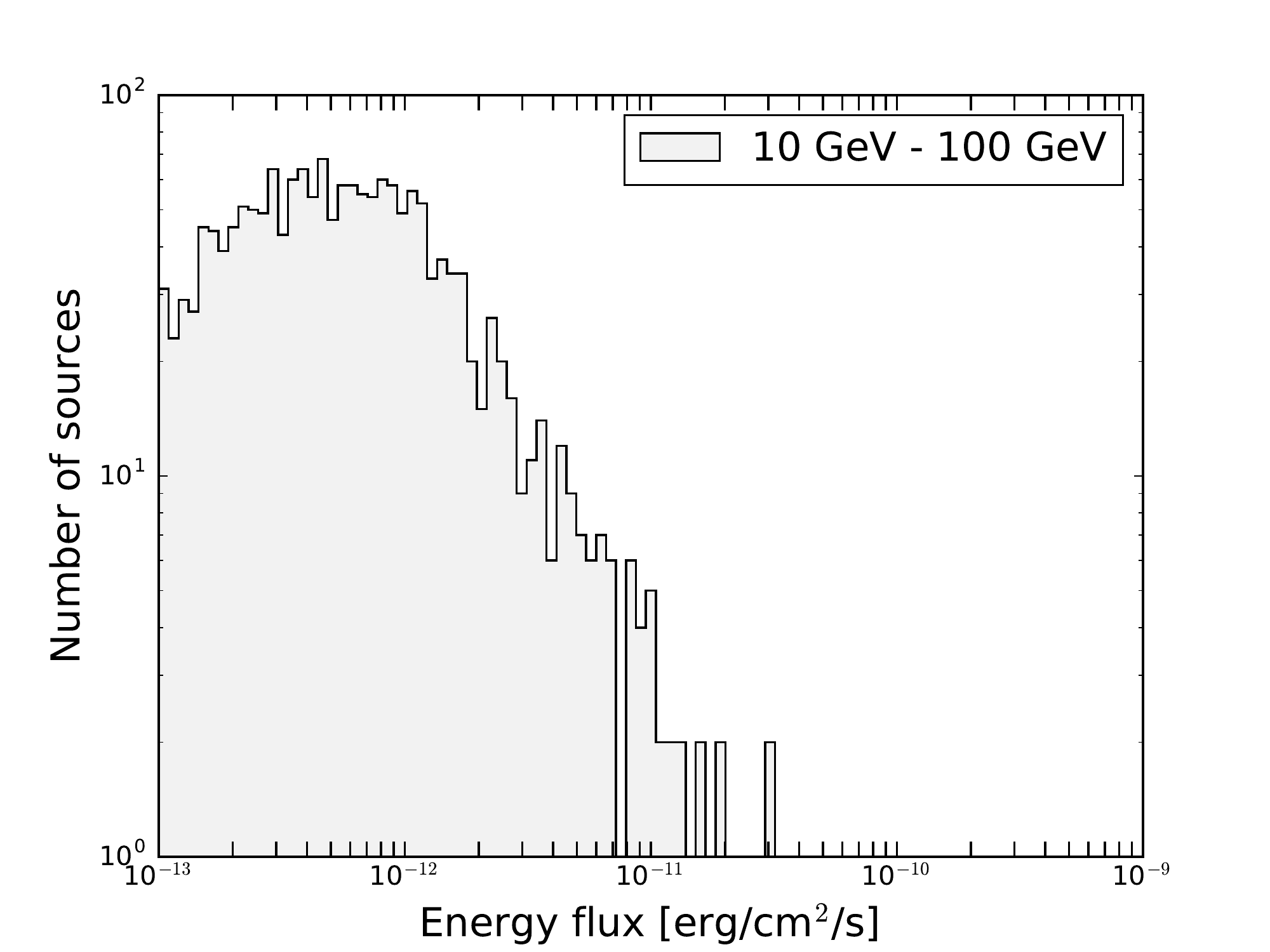}
\caption{Energy flux distribution of sources at high Galactic latitude ($|b| > 10^{\circ}$) in the Fermi LAT 3FGL source catalog, in different photon energy ranges: \emph{Left:} 100 MeV to 100 GeV (blue) and 1 GeV to 100 GeV (green); \emph{Right:} 10 GeV to 100 GeV.\label{efluxhistscombined}}
\end{figure*}

Our aim is to predict the number of dark matter subhalos that might be present as unidentified point sources in the 3FGL catalog. Therefore, we want to define the detectability of a source according to the threshold for inclusion in 3FGL.
In the left panel of Fig.~\ref{efluxhistscombined}, the energy flux distributions of sources in 3FGL at $|b|>10^{\circ}$ are shown for the photon energy range 100 MeV $-$ 100 GeV and for the photon energy range 1 GeV $-$ 100 GeV. The peaks of the distribution are at $\sim 4.0 \cdot 10^{-12} \: \text{erg}/\text{cm}^{2}/\text{s}$ and $\sim 1.35 \cdot 10^{-12} \: \text{erg}/\text{cm}^{2}/\text{s}$, respectively.
The figure also shows that the energy flux above 1 GeV of the faintest source in 3FGL is $\sim 4.0 \cdot 10^{-13} \: \text{erg}/\text{cm}^{2}/\text{s}$.

We expect dark matter annihilation spectra to be harder than the gamma-ray spectra produced by pulsars and blazars, that is, we expect gamma rays from dark matter annihilations to have in general higher energies than gamma rays from blazars or pulsars \cite{fermi3fgl}. Because many unidentified sources are expected to be blazars or pulsars, we choose not to use the detectability threshold inferred from the energy flux distribution in the range \hbox{100 MeV $-$ 100 GeV}, but rather that inferred from the energy flux distribution in the range 1 GeV $-$ 100 GeV. 

As a conservative detectability threshold, we will take the energy flux above 1 GeV corresponding to the peak of the distribution: $F_{\text{cons}} \equiv F_{> 1 \: \text{GeV}} = 1.35 \cdot 10^{-12} \: \text{erg}/\text{cm}^{2}/\text{s}$. We expect the catalog to be complete down to the peak of the distribution, i.e., all sources with fluxes above the peak value have been detected regardless of their spectra. Therefore, we are confident this is a conservative detectability threshold.
As a (very) optimistic detectability threshold, we will take the energy flux above 1 GeV corresponding to the energy flux of the faintest source in 3FGL: $F_{\text{opt}} \equiv F_{> 1 \: \text{GeV}} = 4.0 \cdot 10^{-13} \: \text{erg}/\text{cm}^{2}/\text{s}$.
 
Because the gamma-ray spectrum resulting from dark matter annihilations to $\tau^{+}\tau^{-}$ is harder than that produced in annihilations to $b\bar{b}$, the energy flux in the range 10 $-$ 100 GeV is higher in the case of annihilations to $\tau^{+}\tau^{-}$ than in the case of annihilations to $b\bar{b}$. Therefore, we expect more detectable sources for dark matter annihilating to $\tau^{+}\tau^{-}$ than for annihilations to $b\bar{b}$ if we use a detectability threshold inferred from the energy flux distribution of sources in 3FGL in the energy range 10 $-$ 100 GeV. 
The peak of the energy flux distribution of 3FGL sources in this energy range is \hbox{$4.43 \cdot 10^{-13}  \: \text{erg}/\text{cm}^{2}/\text{s}$} (right panel of Fig. \ref{efluxhistscombined}), and we will use this detectability threshold, denoted by $F_{\text{cons}, \tau^{+}\tau^{-}}$, in the case of annihilations to $\tau^{+}\tau^{-}$.

We exclude all subhalos in VL-II with latitudes smaller than 10 degrees, because a point source at lower latitude --- that is, closer to the Galactic Plane --- would be much harder to detect due to the strong Galactic background and would therefore have to have a higher energy flux than the threshold we inferred from the energy flux distribution of sources in 3FGL at latitudes above 10 degrees.

\section{Results}\label{results-section}
\subsection{Number of detectable subhalos}
The VL-II halo provides a model of the Milky Way, but it cannot capture the orientation of the Galactic disk, much less the specific location of the Earth within this simulated halo. A realistic assumption is that the Galactic disk should be coinciding with the plane perpendicular to the major axis of the dark matter halo. Because the $z$-direction in VL-II is roughly coincident with the major axis of the halo, we assume the Galactic plane to be corresponding to $z = 0$. To generate multiple realisations of the dark matter subhalo population around Earth, we placed the observer at 8.5 kpc (roughly the distance from the Galactic center to Earth) from the center of the halo and rotate it around this point in the $x, y$-plane, keeping the distance to the Galactic center constant. 

\begin{figure*}[!t]
	\centering
		\includegraphics[width=0.48\textwidth]{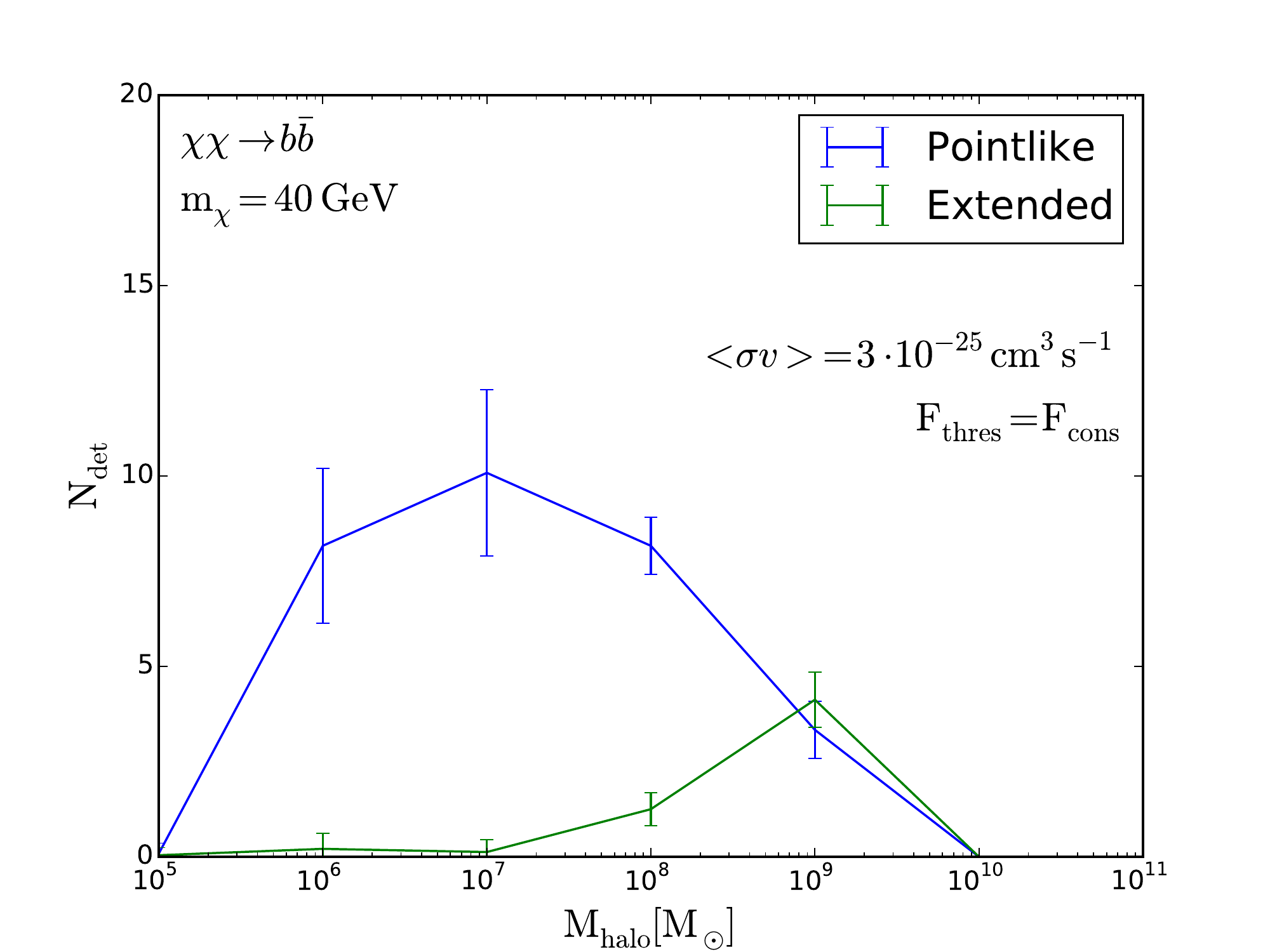}
	     \includegraphics[width=0.48\textwidth]{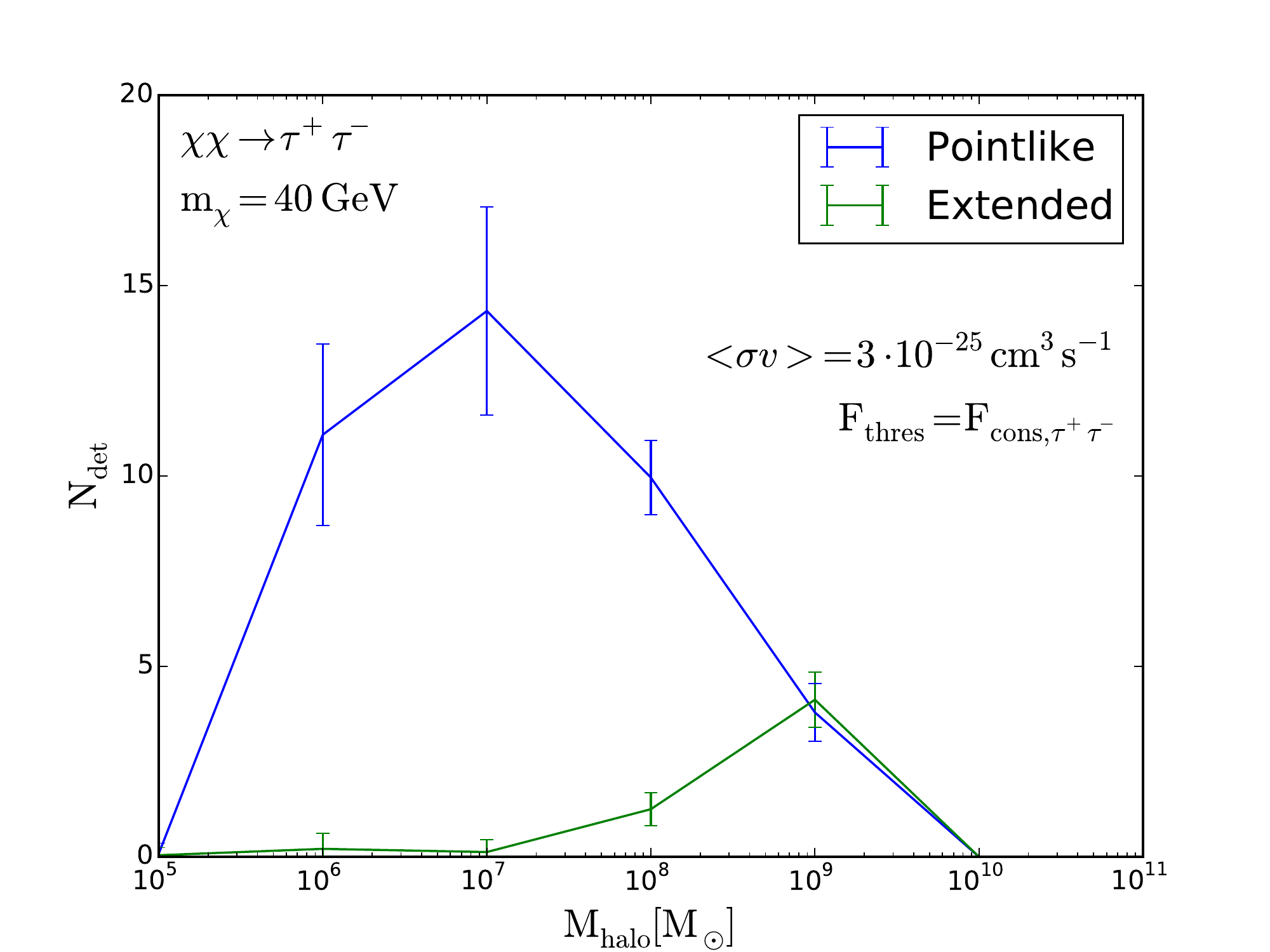}
\caption{Number of detectable point-like (blue) and spatially extended (green) subhalos in VL-II as a function of subhalo mass in the case of a 40 GeV dark matter particle annihilating to $b\bar{b}$ (\emph{left}) or $\tau^{+}\tau^{-}$ (\emph{right}) with a cross section of $\langle\sigma v\rangle \: = 3 \cdot 10^{-25} \: \text{cm}^{3} \: \text{s}^{-1}$ (ten times larger than the thermal cross section). Subhalos are considered detectable if their gamma-ray energy flux above 1 GeV exceeds $F_{\text{cons}}$ (\emph{left}) or if their gamma-ray energy flux above 10 GeV exceeds $F_{\text{cons}, \tau^{+}\tau^{-}}$ (\emph{right}). The error bars correspond to 1 sigma due to rotating the observer position around the Galactic Center. Spatial extension is defined according to Eq. \ref{spatialextension}. The $J$-factors of spatially extended sources were calculated by integrating the luminosity out to to the containment angle of Fermi LAT at $E_{\gamma} = 1$ GeV.  
\label{goedgoed_M_N_40_3*10^(-25)_NFW_planck_tt}}
\end{figure*}

Figure \ref{goedgoed_M_N_40_3*10^(-25)_NFW_planck_tt} shows the number of detectable subhalos as a function of subhalo mass, in the case of annihilations to $b\bar{b}$ and $\tau^{+}\tau^{-}$, respectively. In both plots, a relatively high annihilation cross-section of $3 \cdot 10^{-25} \: \text{cm}^{3} \: \text{s}^{-1}$ is assumed, to illustrate the typical masses of detectable subhalos. Flux thresholds $F_{\text{cons}}$ and $F_{\text{cons}, \tau^{+}\tau^{-}}$ were used for annihilations to  $b\bar{b}$ and $\tau^{+}\tau^{-}$, respectively.
The error bars in the plots correspond to the 1 sigma scatter arising from the 24 different observer locations in the VL-II halo that we used.

\begin{figure*}[!t]
	\centering
		\includegraphics[width=0.48\textwidth]{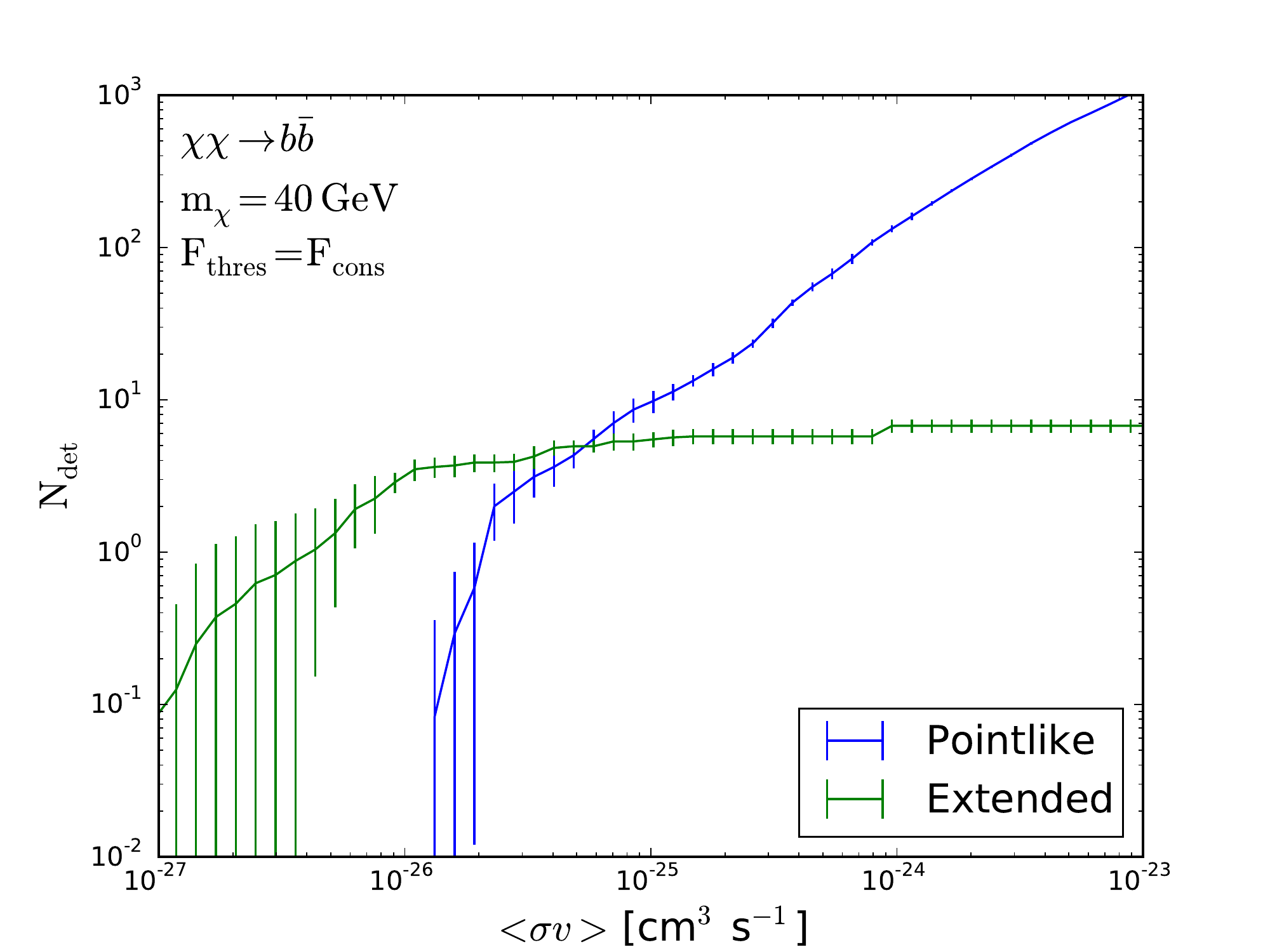}
		\includegraphics[width=0.48\textwidth]{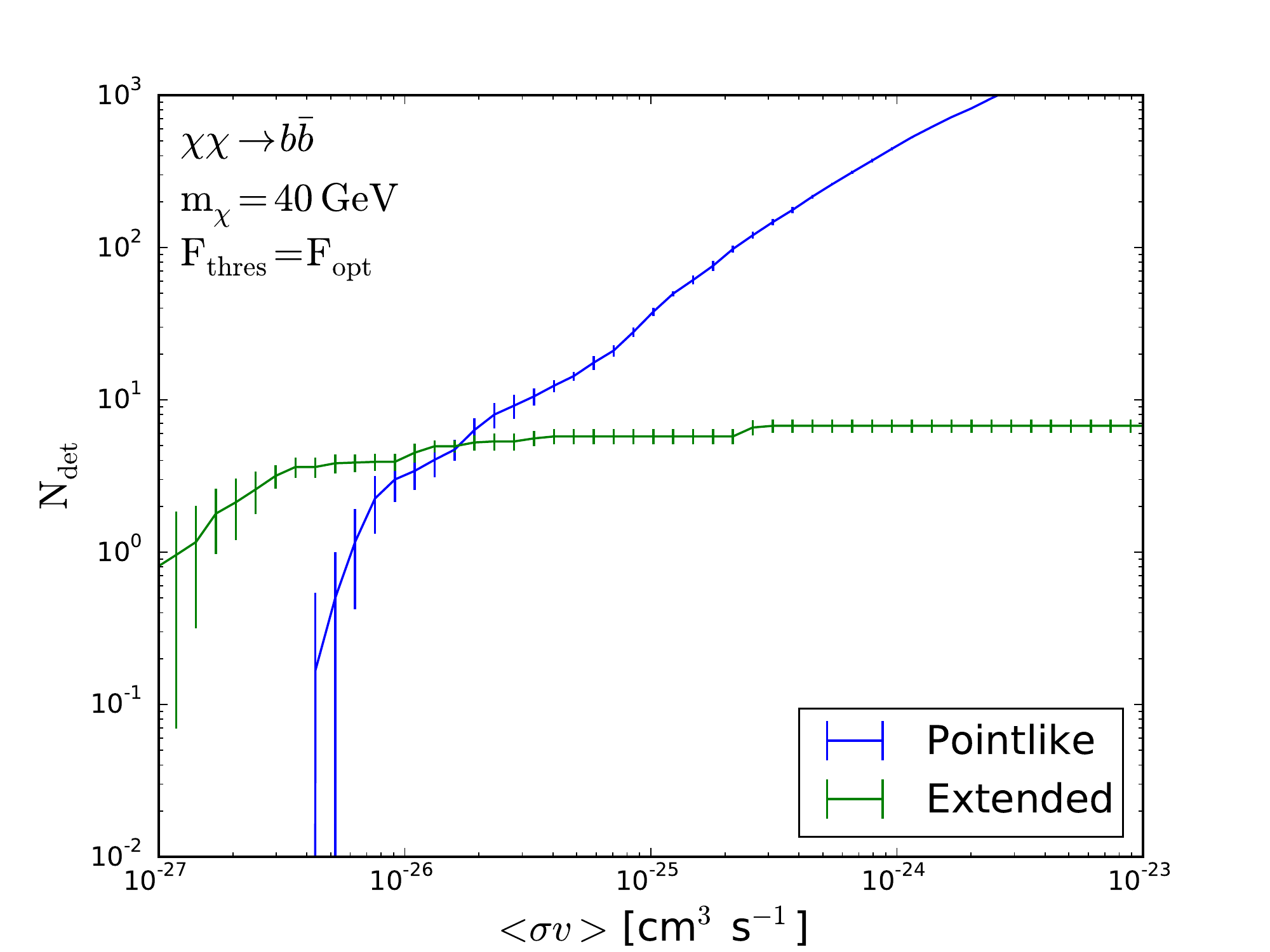}
		\includegraphics[width=0.48\textwidth]{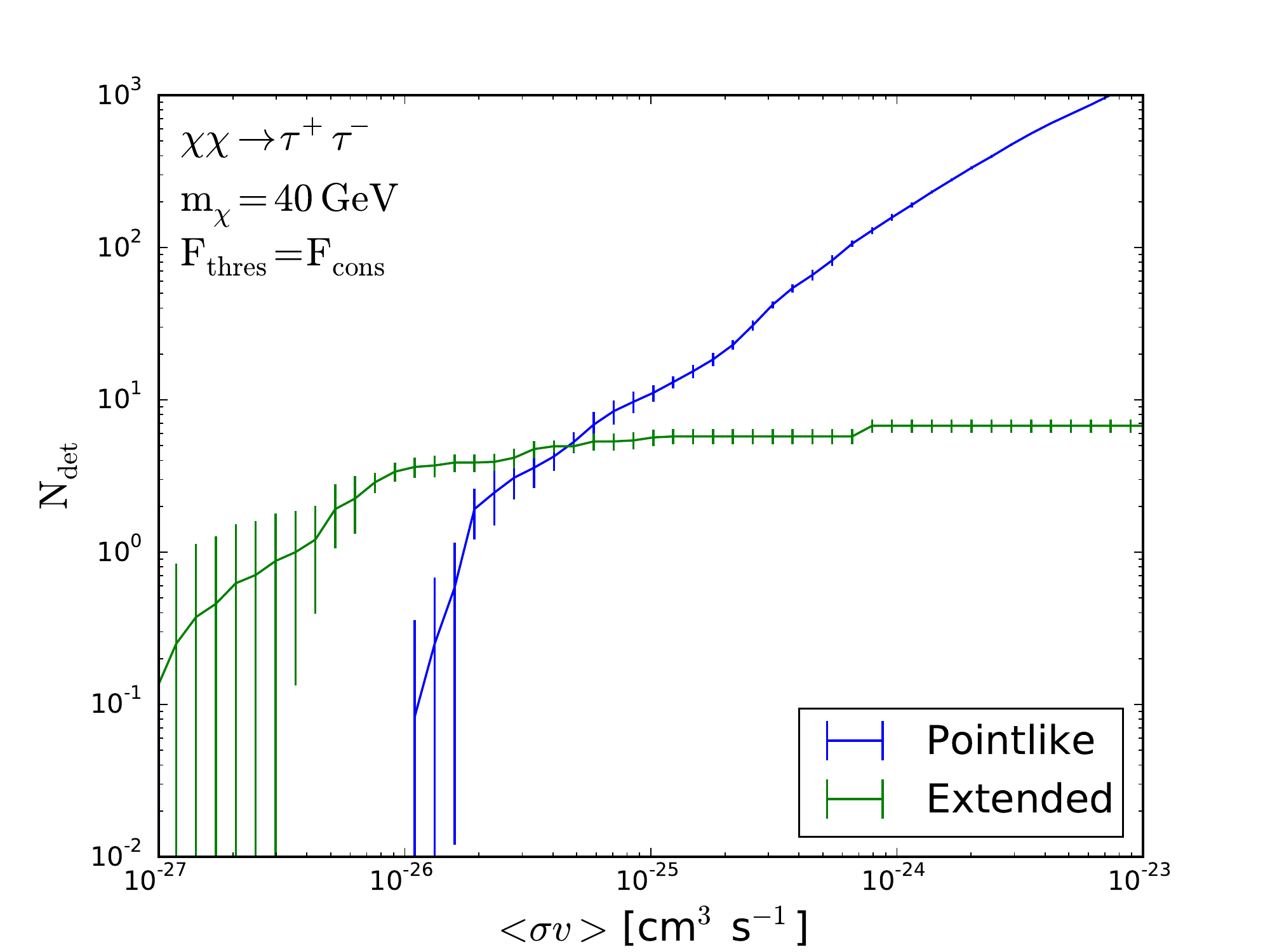}
		\includegraphics[width=0.48\textwidth]{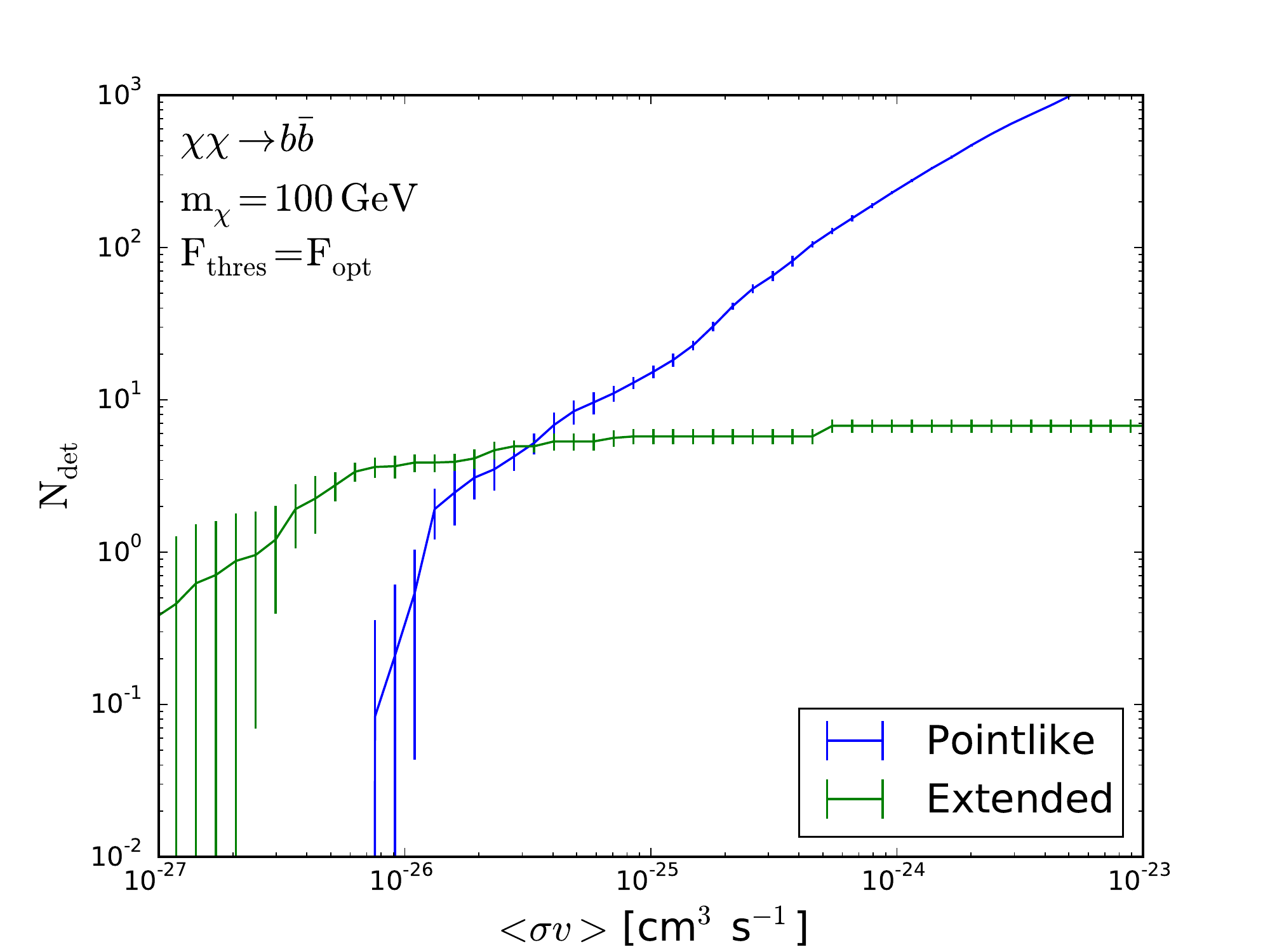}
		\caption{Number of detectable point-like (blue) and spatially extended (green) subhalos plotted against the annihilation cross-section for several scenarios.  The dark matter particle mass, annihilation channel, and  adopted energy flux threshold are indicated on each panel. The error bars correspond to 1 sigma due to rotating the observer position around the Galactic Center. The $J$-factors of spatially extended sources were calculated by integrating the luminosity out to to the containment angle of Fermi LAT at $E_{\gamma} = 1$ GeV. 
\label{paper_N_sigma_100_bb_opt}}
\end{figure*}

\begin{figure*}[t]
	\centering
		\includegraphics[width=0.48\textwidth]{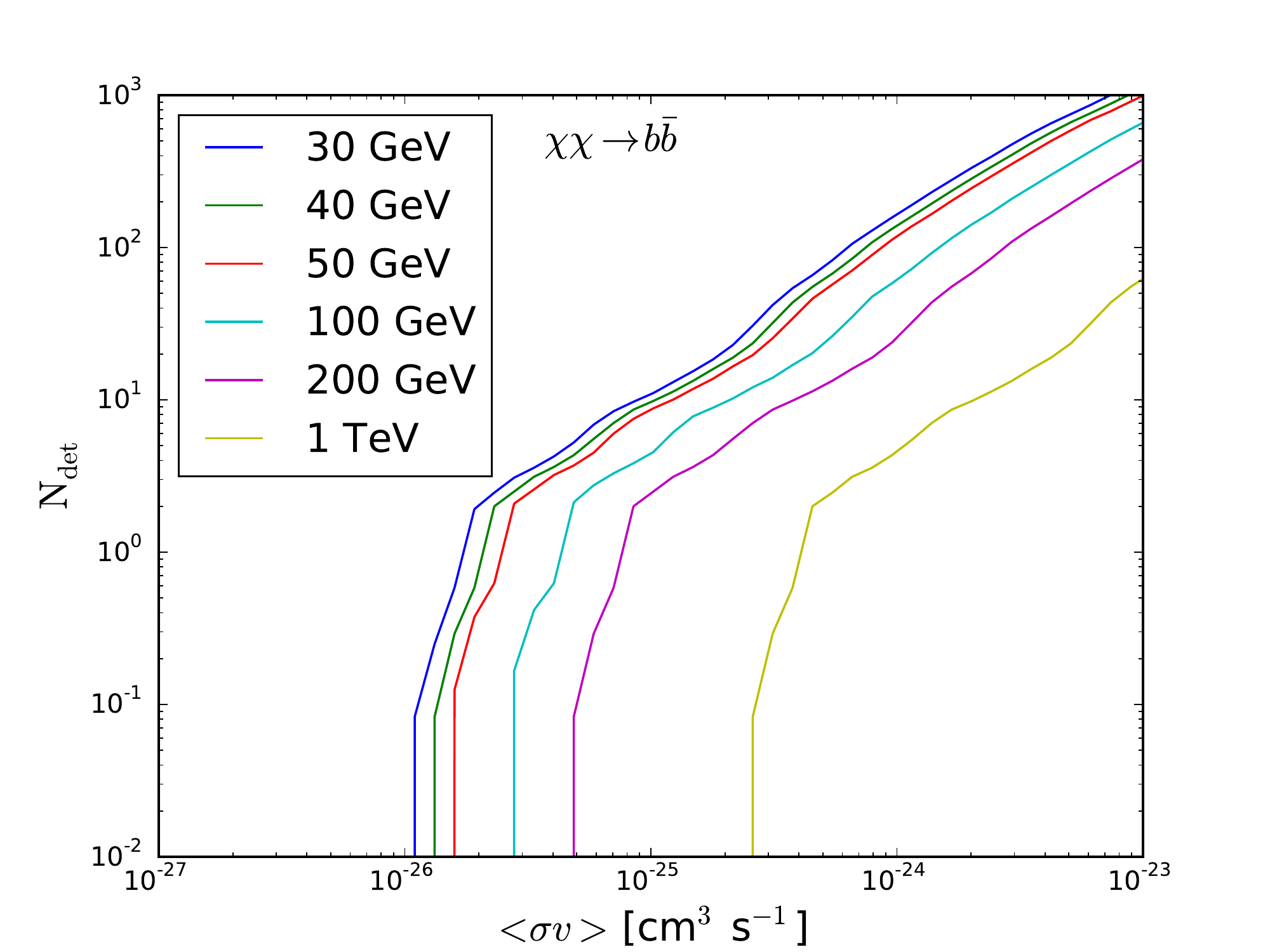}
		\includegraphics[width=0.48\textwidth]{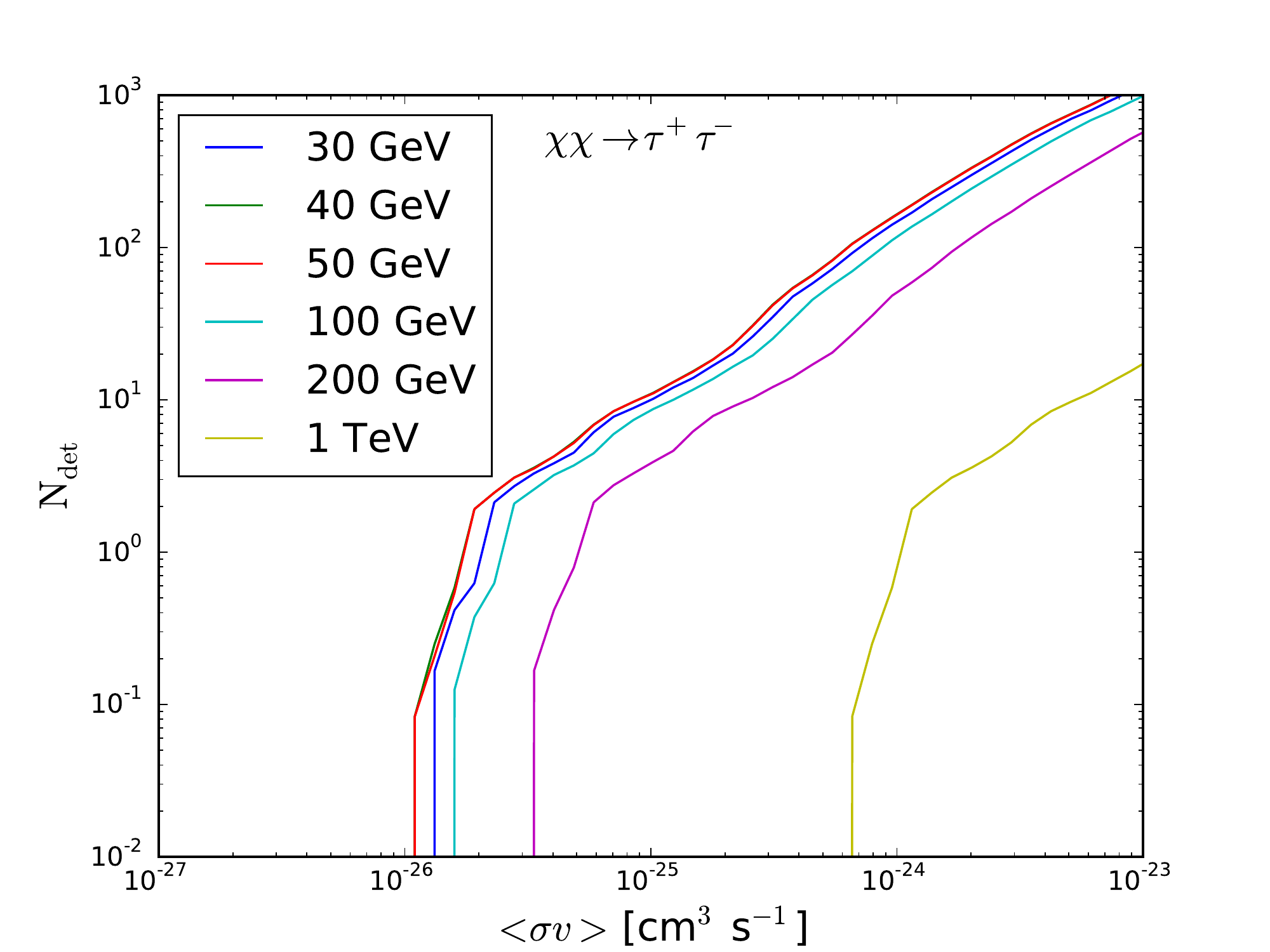}
		\caption{Number of detectable point-like subhalos plotted against the annihilation cross-section for different values of the dark matter particle mass $m_{\chi}$.  The number of detectable subhalos per value of the annihilation cross-section was averaged over 24 observer positions. \emph{Left:} assuming annihilation into $b\bar{b}$, and energy flux threshold $F_{\text{cons}}$. \emph{Right:} assuming annihilation into $\tau^{+}\tau^{-}$, and energy flux threshold $F_{\text{cons}, \tau^{+}\tau^{-}}$.
\label{goedgoed_thesis_N_sigma_bb_NFW_planck_av_allmasses}}
\end{figure*}

\begin{figure*}[t]
	\centering
		\includegraphics[width=0.48\textwidth]{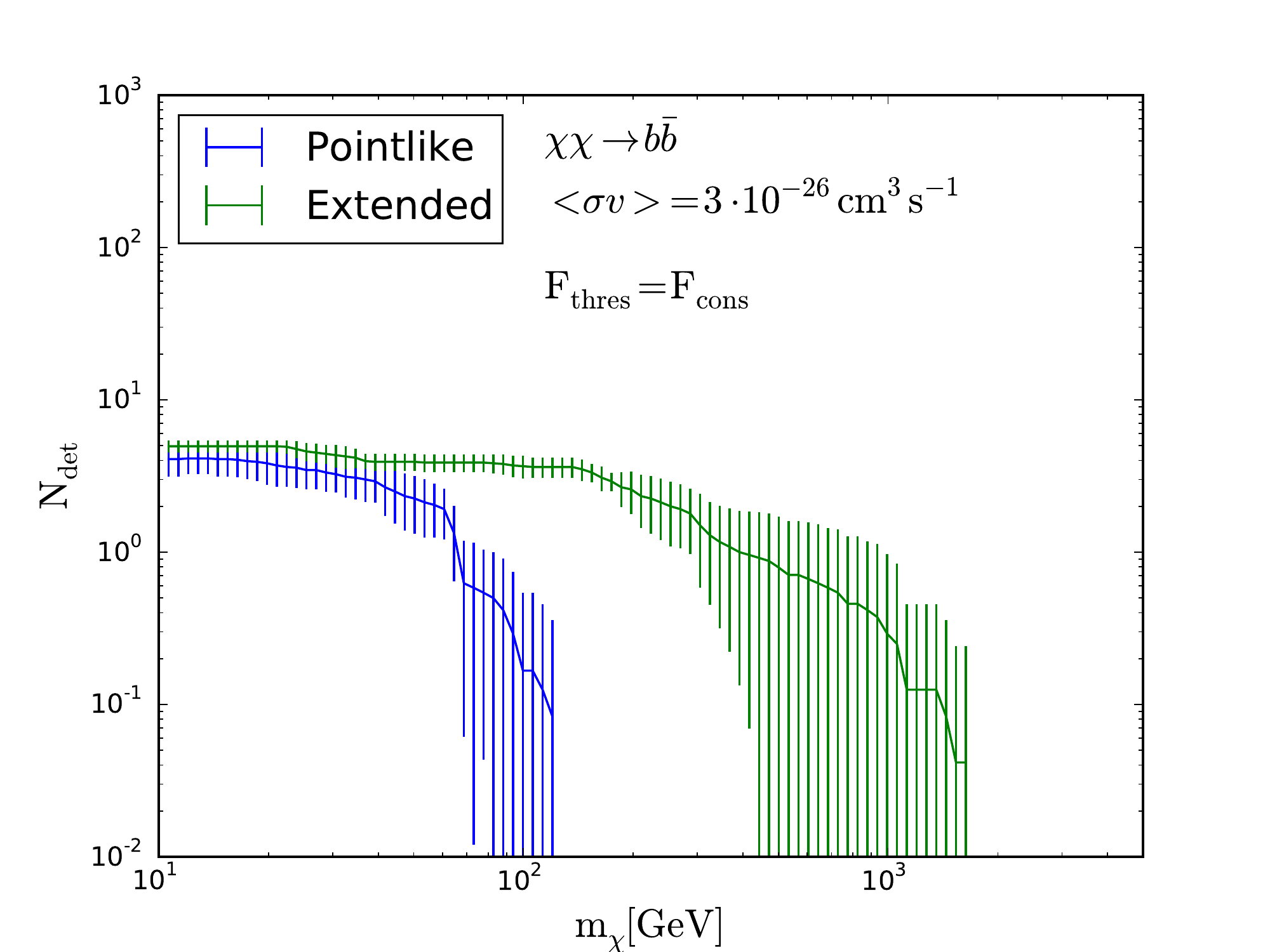}
		\includegraphics[width=0.48\textwidth]{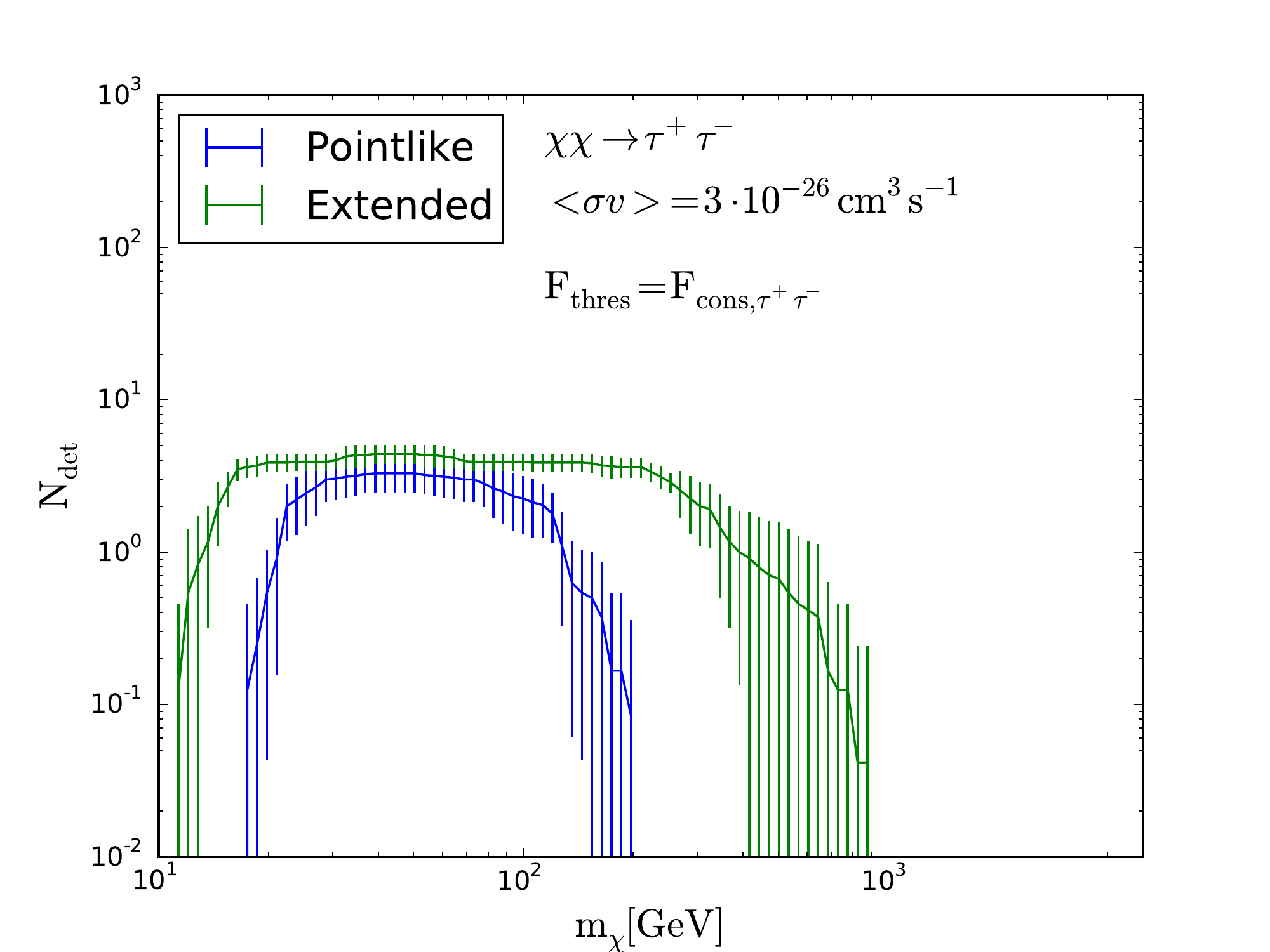}
\caption{Number of detectable point-like (blue) and spatially extended (green) subhalos plotted against dark matter particle mass $m_{\chi}$, for a thermal cross-section and annihilation into $b\bar{b}$ (\emph{left}) and annihilation into $\tau^{+}\tau^{-}$ (\emph{right}). The error bars correspond to 1 sigma due to rotating the observer position around the Galactic Center.\label{paper2_N_mchi_thermal_bbcons}}
\end{figure*}

In Fig.~\ref{paper_N_sigma_100_bb_opt}, the number of detectable subhalos is plotted against the annihilation cross-section for several choices of WIMP masses, annihilation channels and choices of the energy flux threshold. The error bars correspond to 1 sigma due to rotating the observer position around the Galactic Center, like in Fig.~\ref{goedgoed_M_N_40_3*10^(-25)_NFW_planck_tt}, and spatial extension is defined according to Eq. \ref{spatialextension}. The predicted number of detectable subhalos per panel for a thermal cross-section is given in Table \ref{resultstable}.
Fig.~\ref{goedgoed_thesis_N_sigma_bb_NFW_planck_av_allmasses} shows the number of detectable subhalos averaged over observer positions for different choices of the dark matter particle mass in the case of annihilations to $b\bar{b}$ and $\tau^{+}\tau^{-}$, respectively. In Fig.~\ref{paper2_N_mchi_thermal_bbcons} the number of detectable subhalos is plotted against dark matter particle mass for a thermal cross-section and annihilations to $b\bar{b}$ and $\tau^{+}\tau^{-}$, respectively.

\begin{table*}[t]
\centering
\begin{tabular}{c | c | c | c}
$m_{\chi}$ & ann. channel & flux threshold & $\text{N}_{\text{det}}$ \\
\hline
40 & $b\bar{b}$ & $F_{\text{cons}}$ & $2.8 \pm 0.8$ \\
40 & $b\bar{b}$ & $F_{\text{opt}}$ & $9.8 \pm 1.6$ \\
40 & $\tau^{+}\tau^{-}$ & $F_{\text{cons, $\tau^{+}\tau^{-}$}}$ & $3.3 \pm 0.8$ \\
100 & $b\bar{b}$ & $F_{\text{opt}}$ & $4.5 \pm 0.8$ \\
\end{tabular}
\caption{Predicted number of detectable subhalos for different choices of dark matter mass, annihilation channel and flux threshold, in case of a thermal cross-section. Errors correspond to 1 sigma due to 24 different observer positions.}
\label{resultstable}
\end{table*}

\subsection{Upper limits on the annihilation cross-section}
\begin{figure*}[t]
	\centering
		\includegraphics[width=0.48\textwidth]{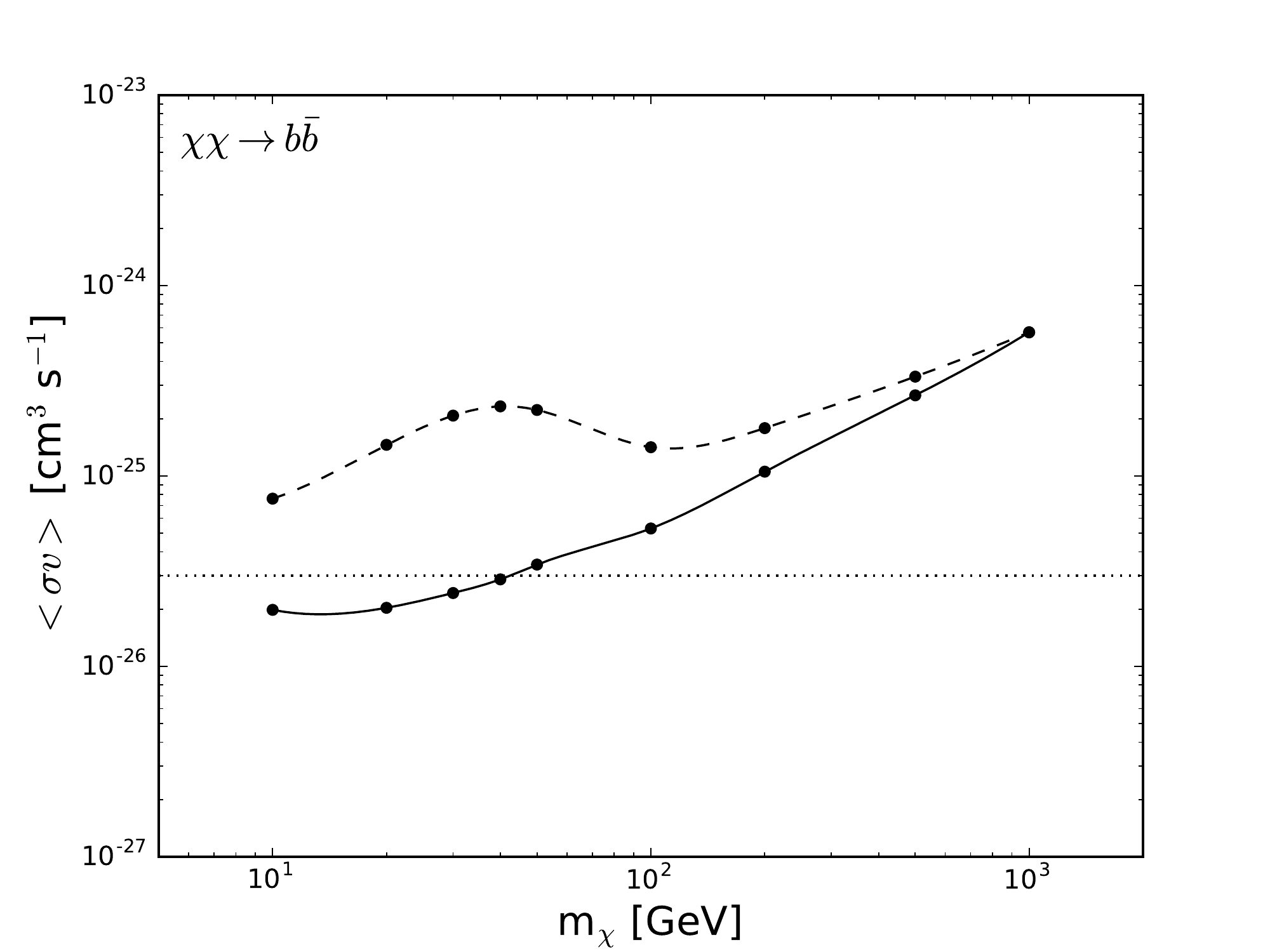}
		\includegraphics[width=0.48\textwidth]{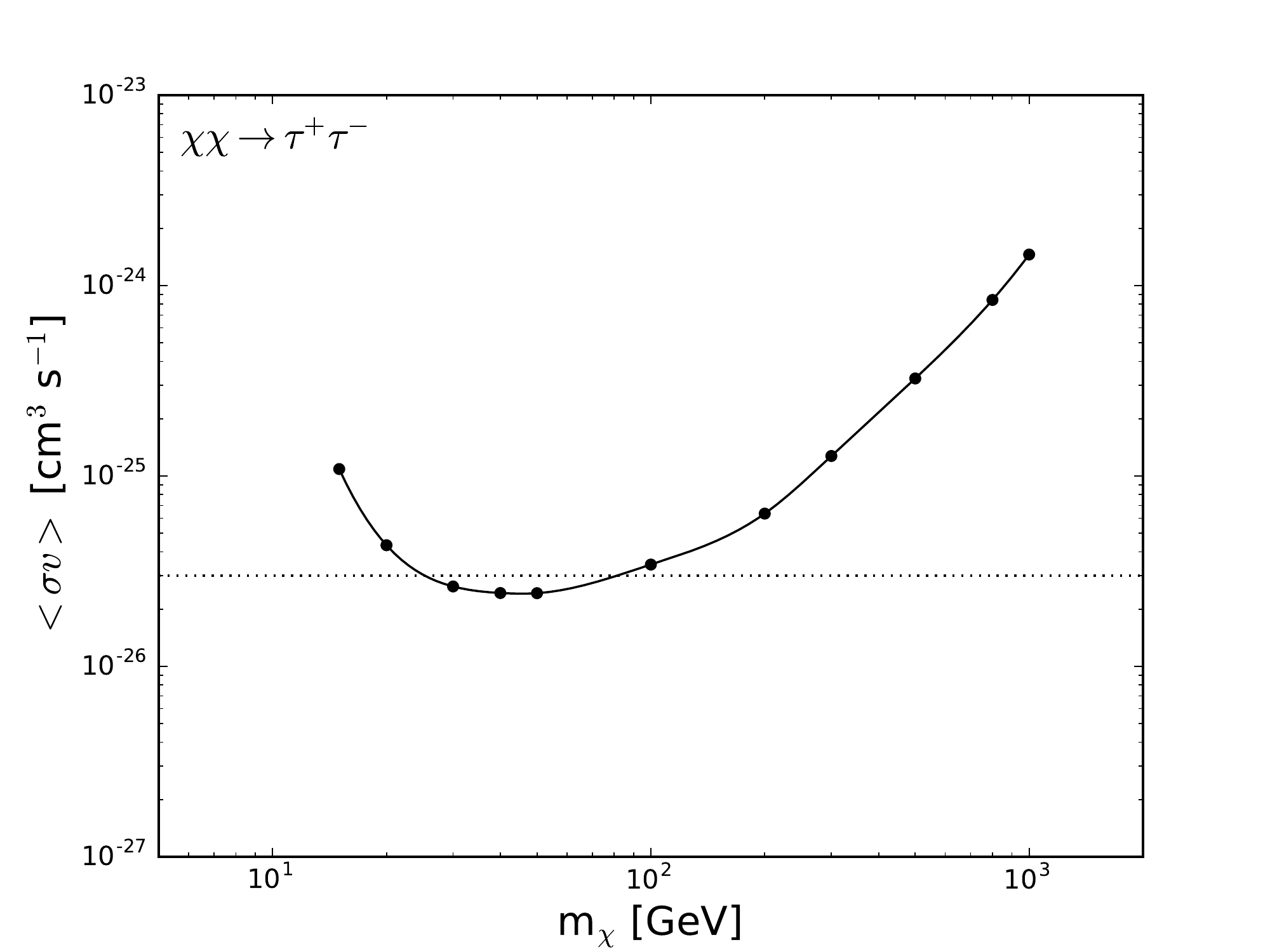}
		\caption{95 \% confidence level upper limits on the dark matter annihilation cross-section as function of the dark matter particle mass $m_{\chi}$, in the case of $100 \%$ annihilation into $b\bar{b}$ (\emph{left}) and $\tau^{+}\tau^{-}$ (\emph{right}). The solid line represents the constraint that would have been obtained if there were no dark matter subhalo candidate sources in 3FGL. The dashed line represents the constraint taking into account the population of candidate sources as provided by \cite{bertoni}. Candidate sources are compatible with dark matter annihilation spectra at 95 \% confidence level. The dotted line corresponds to a thermal cross-section. Subhalos are considered detectable if their gamma-ray energy flux above 1 GeV exceeds $F_{\text{cons}}$ (\emph{left}) or if their gamma-ray energy flux above 10 GeV exceeds $F_{\text{cons}, \tau^{+}\tau^{-}}$ (\emph{right}).\label{upperlimits}}
\end{figure*}

In Fig. \ref{upperlimits} we present our upper limits on the dark matter annihilation cross-section as function of the dark matter particle mass $m_{\chi}$ in the case of annihilations to $b\bar{b}$ and in the case of annihilations to $\tau^{+}\tau^{-}$. The solid lines were obtained by calculating the value of the annihilation cross-section for which 95 \% of the realisations predict one or more detectable subhalos. Thus, the solid line corresponds to our 95 \% confidence level constraints in the case of no candidate dark matter subhalos in 3FGL. For annihilations to $b\bar{b}$ (left panel), the dashed line was obtained in the same way as the solid line, but by requiring that for a given mass $m_{\chi}$, 95 \% of the realisations predict more detectable subhalos than the number of candidate subhalos in 3FGL compatible with that value of $m_{\chi}$, as presented by \cite{bertoni} (see the right panel of Fig. \ref{mdisflux_NFW}). We did not have this information for the case of annihilations to $\tau^{+}\tau^{-}$, and for this scenario we therefore only present the optimistic upper limits on the annihilation cross-section corresponding to the absence of any candidate dark matter subhalos in 3FGL, in the right panel of Fig. \ref{upperlimits}.

\section{Discussion and conclusions}\label{discussion-section}
While other studies \cite[e.g.][]{bertoni} have used a mass function and radial distribution of subhalos in the Milky Way as determined from numerical simulations and then extrapolated to masses below the resolution limits of these simulations, we avoid the uncertainties associated with these extrapolations by restricting our study to halos above the mass resolution limit --- we do not expect the subhalos below the mass resolution limit to be detectable anyway, since such halos would have to be unrealistically close; see the left panel of Fig. \ref{mdisflux_NFW}. Fig. \ref{goedgoed_M_N_40_3*10^(-25)_NFW_planck_tt} confirms this expectation: there are no detectable subhalos in the still well-resolved $10^{5} M_{\odot}$ mass range. Using observable quantities in VL-II to define subhalo density distributions, we have not made assumptions regarding the effect of tidal stripping, as opposed to \cite{bertoni}.

In our results we have also shown the number of spatially extended sources that would be detectable. However, this result should be critically assessed: in calculating this number, we have not performed a line-of-sight integral of the dark matter density squared, rather, we integrated the density squared out to the \hbox{68 \%} containment angle at 1 GeV of Fermi LAT. Moreover, we used a detectability threshold corresponding to the threshold for the inclusion of point sources in 3FGL, whereas spatially extended sources are less detectable than point sources due to the challenges of background modelling \cite{fermi3fgl}.

Adopting the peak of the energy flux distribution above 1 GeV of sources in 3FGL as the energy flux threshold for inclusion in 3FGL, we have found that $2.8 \pm 0.8$ subhalos in a Milky Way-like halo would be detectable with Fermi LAT as a point source at 5-sigma level in the case of a 40 GeV dark matter particle annihilating to $b\bar{b}$-quarks at a thermal cross-section. This is compatible with the findings of \citet{pieri}.
If we instead take an optimistic approach by taking the energy flux above 1 GeV of the faintest source included in 3FGL as the threshold for inclusion in 3FGL, we find $9.8 \pm 1.6$ detectable subhalos.

For a 100 GeV dark matter particle annihilating to $b\bar{b}$ at a thermal cross-section, a scenario that has not yet been excluded by the most stringent upper limits on the annihilation cross-section set by Fermi LAT in their combined dwarf analysis (\cite{fermidwarfs}), we predict $4.5 \pm 0.8$ detectable point-like subhalos using the energy flux above 1 GeV of the faintest source in 3FGL as the threshold for detectability. Although this choice of threshold is probably too optimistic, this result does indicate that even though no significant gamma-ray emission was observed from dwarf galaxies, we might still expect to find dark matter subhalos among the unidentified sources of Fermi LAT.

\begin{figure}[t]
	\centering
		\includegraphics[width=0.5\textwidth]{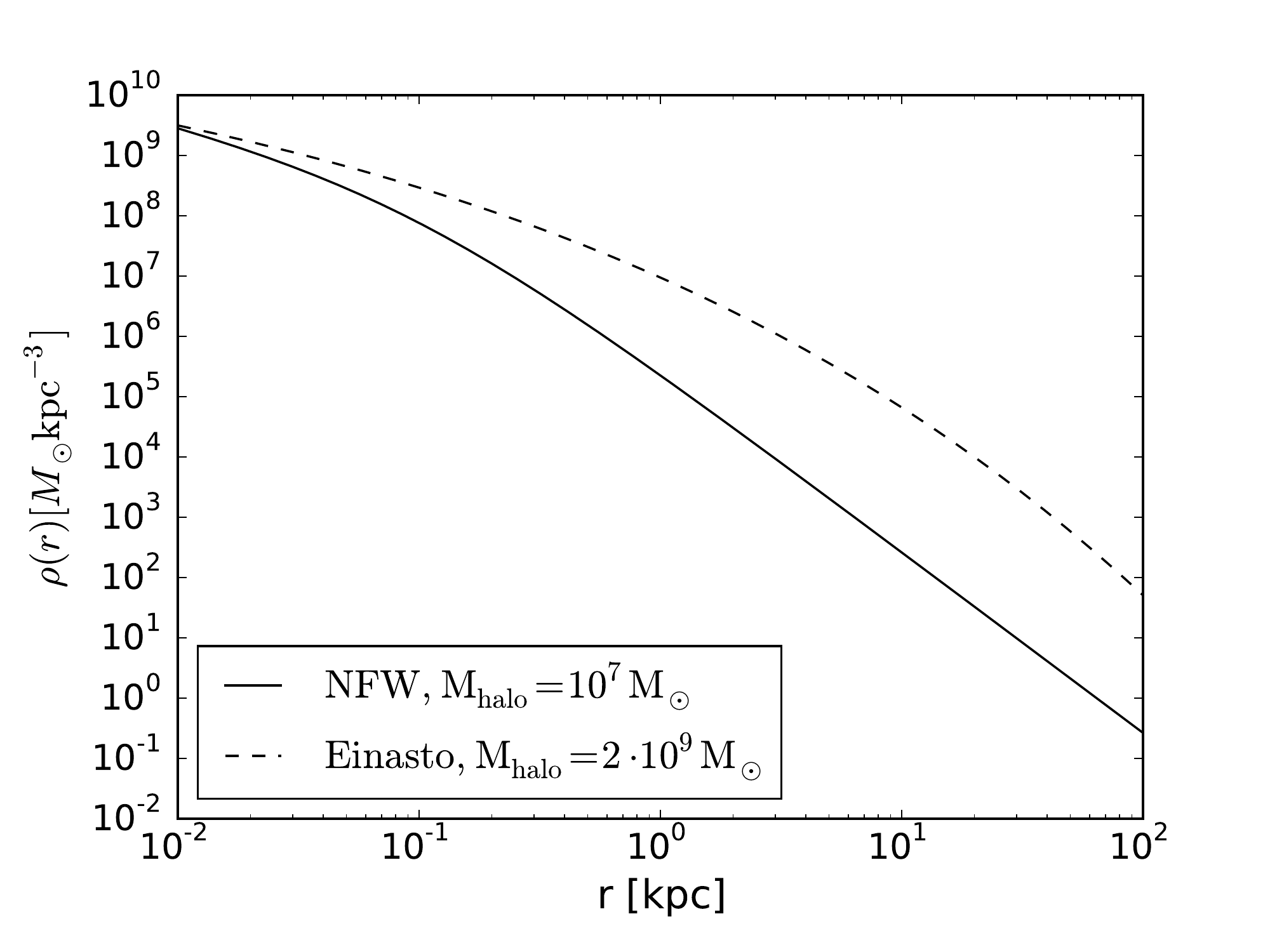}
\caption{Comparison of the NFW profile we assigned to a $10^7 M_{\odot}$ halo in VL-II with the Einasto profile \citet{bertoni} assigned to the same halo before tidal stripping. \label{BHLeinasto}}
\end{figure}

Our results are slightly in tension with the results of \citet{bertoni}. Whereas we predicted $\sim 3$ detectable subhalos for a 40 GeV dark matter particle annihilating to $b\bar{b}$, \citet{bertoni} predicted $\sim$ 10 detectable subhalos for the same dark matter particle parameters and a slightly more conservative detectability threshold. For this reason, our upper limits on the annihilation cross-section using the dark matter subhalo candidate population in 3FGL from \cite{bertoni} are slightly weaker than those presented in \cite{bertoni}. We identify the origin of this discrepancy in their optimistic choice of dark matter density profile. As discussed in Ref. \cite{berlinhooper}, the authors of Ref. \citet{bertoni} describe halos with an Einasto profile, and by comparing the mass fraction in subhalos in the local volume with the mass fraction in subhalos at the virial radius of a halo in the Aquarius simulation, they assume the halos lose 99.5 \% of their initial mass due to tidal stripping, without modifying the density profile within the tidal radius. This means that a $10^7 M_{\odot}$ stripped halo has a scale radius corresponding to a $2 \cdot 10^{9} M_{\odot}$ halo before tidal stripping. In Fig. \ref{BHLeinasto} we compare the density profile of a $10^7 M_{\odot}$ halo used in our analysis with the corresponding profile used in \cite{bertoni}.

If we consider a $10^5 M_{\odot}$ subhalo after tidal stripping, in the analysis of \cite{bertoni} this subhalo had a mass of $2 \cdot 10^7 M_{\odot}$ before infall into the host halo. Following \cite{bertoni} in adopting the concentration-mass relation presented in \cite{sanchezconde}, a $2 \cdot 10^7 M_{\odot}$ halo with a density distribution described by the Einasto profile used by \cite{bertoni} has a scale radius (the radius at which the logarithmic slope of the profile equals -2) of 0.42 kpc (using a NFW profile, the halo would have a scale radius of 0.33 kpc instead).
The radius within which $10^5 M_{\odot}$ is contained is 0.056 kpc. Thus, the authors of Ref. \cite{bertoni} assume that even the parts {\it within} the scale radius of a halo get destroyed due to tidal effects. However, the tidal radii --- defined as the radius at which the density of the subhalo is equal to the density of the host halo \cite{vl2} --- of the subhalos in VL-II are larger than their scale radii. This means that the matter within the scale radii of the subhalos in VL-II have survived tidal stripping, in contradiction with the assumption of \cite{bertoni}.

The main conclusions from our analysis can be summarised as follows:

\begin{enumerate}
\item{} In the case of a 40 GeV WIMP annihilating to $b\bar{b}$ at a thermal cross-section, we predict between $2.8 \pm 0.8$ (using a conservative detectability threshold) and $9.8 \pm 1.6$ (using an optimistic detectability threshold) dark matter subhalos among the unidentified sources of 3FGL. 

\item{} For a 100 GeV WIMP annihilating to $b\bar{b}$ at a thermal cross-section, a scenario that is not excluded by current constraints, we predict $4.5 \pm 0.8$ dark matter subhalos in 3FGL using an optimistic detection threshold.

\item{} We placed upper limits on the dark matter annihilation cross-section using our results obtained for a conservative detectability threshold. These limits are competitive with constraints by other studies in the case that there are no dark matter subhalo candidates in 3FGL.

\item{} Our results are in tension with those of \citet{bertoni}, due to their overly optimistic assumptions concerning the amount of tidal stripping.

\end{enumerate}

Note added: As we were finalising our work, a new analysis by \citet{zhu} showed that the stellar disk of a Milky Way-size galaxy may substantially deplete subhalos near the central region. The total number of low-mass subhalos in their hydrodynamic simulation is roughly half the number that are in a dark-matter--only simulation. If confirmed, this effect would further reduce the detectability of dark matter subhalos.

\section*{Acknowledgements}
We would like to thank Christoph Weniger, Mark Lovell, and Arianna di Cintio for useful discussions, and Francesca Calore, Dan Hooper, and Christoph Weniger for comments on the manuscript. J.G. acknowledges support from a Marie Curie International Incoming Fellowship in the project ``IGMultiWave'' (PIIF-GA- 2013-628997).  G.B. acknowledges support from the European Research
Council through the ERC starting grant WIMPs Kairos. J.D. was supported by the Swiss National Science Foundation.

\appendix
\section{Estimating subhalo parameters from the VL-II simulation }
\subsection{From $r_{Vmax}$ and $V_{max}$ to NFW profile parameters}
Two observable quantities of subhalos in the VL-II simulation are the maximum circular velocity, $V_{max}$, and the radius at which this velocity is reached, $r_{Vmax}$. For a NFW density profile, the following relations hold
\cite{diemand2007}
\begin{equation}\label{rs}
r_s = r_{Vmax} / 2.163
\end{equation}
\begin{equation}\label{vc}
V_{c}^{2}(r) = 4 \pi G \rho_{s} r_{s}^3 \frac{f(r)}{r},
\end{equation}
where
\begin{equation}
f(r) = \ln \left(1 + \frac{r}{r_s} \right) - \frac{r/r_{s}}{1+r/r_{s}}.
\end{equation}
Plugging in $V_{max}$ for $V_{c} (r)$ in Eq. \ref{vc}, we find
\begin{equation}
V_{max}^{2} = 4 \pi G \rho_{s} r_{s}^{3} \frac{1}{2.163 r_{s}} \left[ \ln(3.163) - \frac{2.163}{3.163} \right],
\end{equation}
such that we can write $\rho_{s}$ in terms of $r_{s}$ and $V_{max}$:
\begin{equation}\label{rhos}
\rho_{s} = \frac{2.163 \cdot V_{max}^{2}}{4 \pi G r_{s}^2 \: [ \ln(3.163) - 2.163 / 3.163]}
\end{equation}
Using Eqs. \ref{rs} and \ref{rhos}, we can fully determine a NFW profile for each halo in the simulation. The virial radius $r_{200}$
can be found by requiring that 200 times the critical density equals the average density within the virial radius.

\subsection{Scaling $r_{\text{Vmax}}$ to Planck15 cosmological parameters}
The concentrations and velocity profiles of subhalos in CDM simulations such as VL-II are dependent on the adopted cosmological parameters \cite{ricotti}. As predicted by the theory of hierarchical structure formation, the later small-mass halos (which could eventually become subhalos in a Milky Way-like host halo) form in the history of the Universe, the less concentrated they are, reflecting the lower density of the Universe at later times. Therefore, adopting cosmological values that shift the small-mass halo formation to later epochs results in less concentrated subhalos. The concentration of a halo is related to its radius of maximum circular velocity $r_{\text{Vmax}}$. In \cite{ricotti}, Polisensky and Ricotti show how $r_{\text{Vmax}}$ in CDM simulations scales with the cosmological parameters $\sigma_{8}$ and $n_{s}$ at fixed $V_{max}$ (they found no dependence of $r_{\text{Vmax}}$ on other cosmological parameters). This scaling is given by:

\begin{equation}
r_{\text{Vmax}} \propto (\sigma_{8} 5.5^{n_{s}})^{-1.5}
\end{equation}
The VL-II simulation is based on the WMAP3 cosmology and has $\sigma_{8}$ = 0.74 and $n_{s}$ = 0.951. The latest Planck results provide $\sigma_{8}$ = 0.82 and $n_{s}$ = 0.9667 \cite{planck2015}, such that the scaling becomes:

\begin{equation}
r_{\text{Vmax, Planck15}} = \frac{r_{\text{Vmax, VLII}}}{1.21}\label{rvmaxscaling}
\end{equation}
In our analysis, we used Eq. \ref{rvmaxscaling} to scale the $r_{\text{Vmax}}$-values of the halos in VL-II to the latest values of the cosmological parameters as found by the Planck satellite. For a given maximum circular velocity, the radius of maximum circular velocity becomes smaller, which means that the subhalos in VL-II become more concentrated after applying Eq. \ref{rvmaxscaling}.

\bibliography{Jfactors}

\end{document}